\documentclass[aps,prl,showpacs,showkeys,twocolumn]{revtex4-1}
\usepackage{graphicx,hyperref,amsmath,amsfonts}
\usepackage{epstopdf,color,bm,multirow,rotating,soul}

\begin{document}

\title[Programmable quantum motherboard for logical qubits]{Programmable quantum motherboard for logical qubits}
\author{Nikolay Sergeevich Perminov$^{1,2}$, Diana Yurevna Tarankova$^{3}$, and Sergey Andreevich Moiseev$^{1,2,*}$}
\affiliation{$^{1}$ Kazan Quantum Center, Kazan National Research Technical University n.a. A.N.Tupolev-KAI, 10 K. Marx, Kazan 420111, Russia}
\affiliation{$^{2}$ Zavoisky Physical-Technical Institute, Kazan Scientific Center of the Russian Academy of Sciences, 10/7 Sibirsky Tract, Kazan 420029, Russia}
\affiliation{$^{3}$ Department of Radio-Electronics and Information-Measuring Technique, Kazan National Research Technical University n.a. A.N.Tupolev-KAI, 10 K. Marx, Kazan 420111, Russia}
\email{s.a.moiseev@kazanqc.org}
\pacs{03.67.-a, 03.67.Hk, 03.67.Ac, 84.40.Az}
\keywords{quantum motherboard, logical qubits and qutrits, high-Q resonators, quantum interface,  periodic frequency comb, spectrum optimization.}

\begin{abstract}
We propose a scheme of a programmable quantum motherboard based on the system of three interacting high-Q resonators coupled with two-level atoms. By using the algebraic methods, we found that the investigated atomic-resonator platform can possess an equidistant spectrum of the eigenfrequencies at which simple reversible dynamics of the single photon excitation becomes possible. It was shown that such multiresonator quantum motherboard scheme allows to achieve efficient and programmable quantum state transfer between distributed atoms and generate logical qubits and qutrits. The use of the  studied circuit in quantum processing is proposed.
\end{abstract}

\maketitle

\section{Introduction}
The creation of an effective quantum motherboard (QMB) with long-lived subsystems is of critical importance for quantum information technologies \cite{Hammerer2010,Kurizki2015}.
In practical implementation of multiqubit QMB, it is required a sufficiently strong and reversible interaction of light/microwave qubits with many long-lived quantum systems \cite{Roy2017}, in particular with NV-centers in diamond \cite{Jiang2009}, rare-earth ions in inorganic crystals \cite{Zhong2015} and quantum dots  \cite{Zhang2019}.
In this approach, the best realization of the controllable qubit transfer between distant nodes  provided quantum  efficiency up to 92 \% \cite{Cho2016,Hsiao2018}, while practical quantum computers require at least 99.9 \%.

Robust solution of the high efficiency transfer could be possible with the proper use of the rich features of the many-particle dynamics \cite{Hartmann2008,Hur2016,Noh2017}.
Herein, the basic problem is a construction of the multi-particle systems \cite{Mirhosseini2019cavity} providing convenient control of the time-reversible dynamics. 
The considerable progress in construction of high-Q resonators \cite{Gorodetsky1999,Vahala2003,Kobe2017,Toth2017,Megrant2012} with convenient spatial control opens rich spectral opportunities for the solution of this problem in a system of coupled resonators \cite{Armani2003,Liu2018,Xie2018,Flurin2015,Pfaff2017,Sirois2017}.
In particular, the system of resonators with periodic frequencies coupled to the common waveguide \cite{EMoiseev2017,Moiseev_2017_PRA,Moiseev2018} can extend the dynamical capabilities of linear chains of resonators with the same frequencies \cite{Heebner2002slow}.
Moreover, such multiresonator schemes demonstrate a significant increase in the operating spectral range of the quantum interface (QI). 
In these systems, it is possible to considerable enhance the coupling constant of light signals with resonant atoms herewith the broadband system of the high-Q resonators allows reducing the effects of relaxation and decoherence due to the transition to faster storage processes. 
These properties promise getting higher QI efficiency and using these systems \cite{Perminov2018superefficient,Perminov2019spectral} in decoherence free quantum processing \cite{Kurizki2015,Kockum2018}.

Multresonator system with atoms and controlled coupling between resonators seems useful as efficient tool for entanglement generators, quantum memory,  interface \cite{Flurin2015,Moiseev2018} and it can also be applicable for construction photon-photon multi-qubit gates \cite{Moiseev2013quantum,Andrianov2019cnot}.
In this work, we show that combining the system of high-Q resonators with atoms (quantum dots, NV-centers or long-lived spin systems) situated in these resonators could be a realible approach to obtain full-fledged quantum platform for quantum computation with different type of logical qubits.

\begin{figure}[t]
\includegraphics[width = 0.45\textwidth]{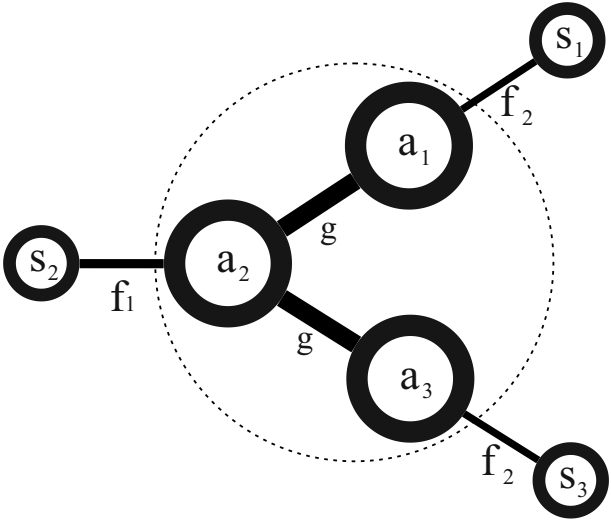}
\caption{
Principle scheme of a programmable quantum motherboard: large circles -- resonators connected by a dynamically controlled coupling $g$, small circles -- two-level atoms geometrically located inside resonators. The dashed line is the geometric border of the board.}
\label{Scheme}
\end{figure}

The proposed platform consists of 3 high-Q resonators forming a controllable frequency comb structure and interacting with long-lived resonant atoms (Fig.~\ref{Scheme}).
A small number of the coupled resonators greatly facilitates the spectral-topological optimization of parameters of the QMB \cite{Perminov2019spectral}.
We analytically and numerically study the optimization criterion  of the platform to obtain reversible and controlled atomic dynamics.
By using the optimization method, we have found the optimal values of free used parameters of the QMB scheme and show that achieving the super high efficiency of the interatomic quantum state transport takes place at the equidistant  eigenfrequencies of the studied platform.
Finally we discuss the obtained results, its possible experimental implementation and potential application.

\section{Theoretical model}
The basic idea of the scheme under consideration is based on the photon echo quantum memory approach demonstrating reversible dynamics in the inhomogeneously broadened quantum systems \cite{Moiseev2004,Tittel2009} especially on its variant of the periodic spectral structure of the inhomogeneous broadening known as the AFC protocol \cite{Riedmatten2008,Akhmedzhanov2016}, and its realization in the optimal resonator \cite{Moiseev2010,Afzelius2010,2013-Sabooni-PRL}. Moreover this approach has been also recently extended to the systems of several coupled resonators \cite{EMoiseev2017,Moiseev_2017_PRA,Perminov2018superefficient}.

We analyze the dynamics of $3$ single mode resonators and resonant two-level systems by using the quantum optics approach  \cite{Walls}. For the case of single-photon excitation considered in this paper, the total quantum state is 
$|\Psi(t)\rangle=e^{-i\omega_0t}\{\sum_{n} s_n(t)|1_{n}^{atom};0^{res}\rangle+ \sum_{n}a_m(t) |0^{atom};1_{m}^{res}\rangle\}$ (where $|0^{atom}\rangle$ and $|0^{res}\rangle$ are the ground states of atoms and cavity modes, $|1_n^{atom}\rangle$ and $|1_m^{res}\rangle$ are exited states of n-th atom and m-th cavity mode with ground state of other quantum systems).
In the framework of this approach, we get the system of linear equations for the slowly varied atomic coherences $s_{n}(t)$ and  field mode amplitudes $a_n(t)$:
\begin{align}\label{eq}
& \nonumber [\partial_{t}-i\Delta]s_1(t)+if_2a_1(t)=0, \\
& \nonumber \partial_{t}s_2(t)+if_1a_2(t)=0, \\
& \nonumber [\partial_{t}+i\Delta]s_3(t)+if_2a_3(t)=0, \\
& \nonumber [\partial_{t}-i\Delta]a_1(t)+if_2s_1(t)+iga_2(t)=0, \\
& \nonumber \partial_{t}a_2(t)+if_1s_2(t)+iga_1(t)+iga_3(t)=0, \\
& {[}\partial_{t}+i\Delta{]}a_3(t)+if_2s_3(t)+iga_2(t)=0,
\end{align}
where the full atomic and field modes are obtained by multiplying the slow modes by the factor $e^{-i\omega_0t}$, $\omega_0$ is a central frequency, $\Delta$ is the frequency detuning of the 1st and 3rd resonators relative to the second one, $g$ is the interaction constant between the resonators, $\{f_1,f_2\}$ is the interaction constant between atoms and field modes and the resonant frequencies of the atoms are equal to the resonant frequencies of the resonators in which they are located.
We also ignored the relaxation terms  \cite{Scully,Walls} in the equation (\ref{eq}), focusing only on the searching for the fast quantum transfer in the QMB scheme with weak decoherence processes in the resonators and atomic qubits.

\section{Laplace solution and spectra}
By assuming that initially  2-nd atom is in the excited state ($s_2(t=0)=1$) and applying the Laplace transform of (\ref{eq}), we obtain the system of the algebraic equations: 
\begin{align}\label{eq_lapl}
&\nonumber [p-i\Delta]s_1+if_2a_1=0, \\
&\nonumber ps_2-1+if_1a_2=0, \\
&\nonumber [p+i\Delta]s_3+if_2a_3=0, \\
&\nonumber [p-i\Delta]a_1+if_2s_1+iga_2=0,\\
&\nonumber pa_2+if_1s_2+iga_1+iga_3=0, \\
&{[}p+i\Delta{]}a_3+if_2s_3+iga_2=0,
\end{align}
where for all the field modes the Laplace transform is defined as $u(p)=\int_0^{\infty} dp~e^{-p t}u(t)$, where $\omega=ip$ is the frequency counted from the central frequency of the radiation $\omega_0$.

We find the solution of (\ref{eq_lapl}) for the amplitude $s_2(p)$:
\begin{align}\label{sol_eq}
& \nonumber s_2=Det^{-1}p(p^4+2(\Delta^2+g^2+f_2^2)p^2 \\
&\nonumber +\Delta^4+2\Delta^2g^2-2\Delta^2f_2^2+2g^2f_2^2+f_2^4),  \\
&\nonumber Det=p^6+(2\Delta^2+2g^2+f_1^2+2f_2^2)p^4+ \\
&\nonumber(\Delta^4+
2(g^2+f_1^2-f_2^2)\Delta^2+
2(g^2+f_1^2)f_2^2+f_2^4)p^2 \\
&+f_1^2(\Delta-f_2)^2(\Delta+f_2)^2,
\end{align}
where $Det$ is the determinant of the linear algebraic equation (\ref{eq_lapl}), which determines the eigenvalues $p_n=-i\omega_n$ according to the standard rule $Det(p=p_n)=0$.

\begin{figure}[t]
\includegraphics[width = 0.45\textwidth]{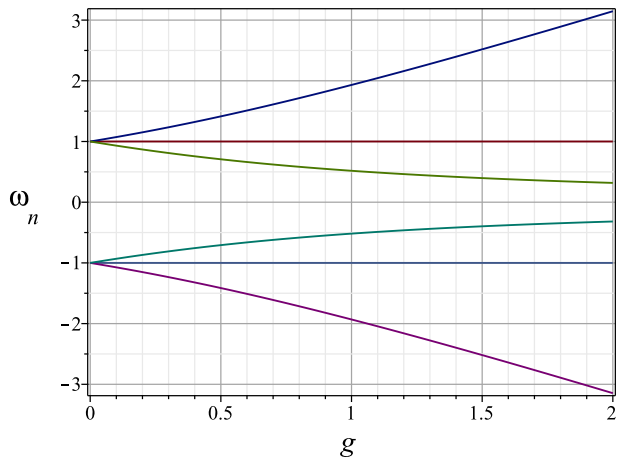}
\caption{Eigenfrequencies $\omega_n$ versus $g$ for $f_1=f_2=1, \Delta=0$.}
\label{spectra_non_optimize}
\end{figure}

In Fig.\ref{spectra_non_optimize} we see a typical dependence of the eigenfrequency distribution of our system versus  $g$ for the case when the initial eigenfrequencies (in the absence of interaction) and coupling constants of all resonators and atoms are the same ($f_1=f_2=1, \Delta=0$).
We see that the initial spectrum (without interaction), consisting of only one frequency, is split into exactly 6 different frequencies (3 independent frequency pairs) due to the interaction between atoms and resonator modes.
For such a situation, all 3 independent frequencies are controlled by only one free parameter $g$, but it requires to use at least 3 free parameters to get the full control at three frequencies.

\section{Equidistance criterion for eigenfrequencies}
The appearance of reversible/periodic dynamics in a closed multiparticle system is possible only if the multiplicity of eigenfrequencies determined by the determinant of system (\ref{sol_eq}).
One of the case for implementing the condition of frequency multiplicity is the condition for the presence of an equidistant frequency comb, which was previously used to implement a highly efficient QI and quantum memory \cite{Moiseev2018,Perminov2019spectral}.
Equidistance criterion for a 6-particle system with eigenfrequencies $\omega_{\pm1},\omega_{\pm2},\omega_{\pm3}$ ($\omega_1\leq\omega_2\leq\omega_3$, $\omega_{-n}=-\omega_n$) without degenerate levels is $\omega_1:\omega_2:\omega_3=1:3:5$ and corresponds to the fulfillment of the condition $\delta=0$, where the non-equidistance error 
\begin{align}\label{err}
\delta=|\omega_2/\omega_1-3|+|\omega_3/\omega_1-5|.
\end{align}

\begin{figure}[t]
\includegraphics[width = 0.45\textwidth]{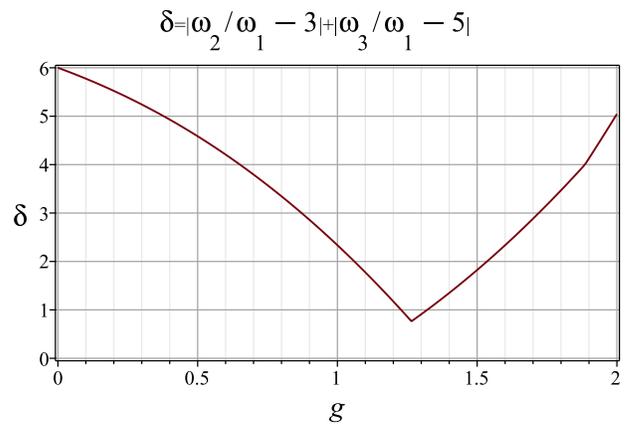}
\caption{Error non-equidistance $\delta$ versus $g$.}
\label{error_eq}
\end{figure}

Firstly we analysed the case of resonant interaction: $f_1=f_2=f=1, \Delta=0$. Using the algebraic solution (\ref{sol_eq}), we didn't find any efficient transfer of the initial 2-nd atomic excitation to the 1-st and 3-nd atoms for arbitrary value $g$ at any moment of time.  
We get spectra of the atom-resonator systems (called also as atomic-photon molecule) depicted in Fig.\ref{spectra_non_optimize} and its non-equidistance error (\ref{err}) shown in Fig.\ref{error_eq}.
Is is seen in Fig.\ref{error_eq}, it is impossible to obtain an equidistant frequency comb ($\delta=0$) for the used parameters for any value of $g$. So, we draw our attention to the nonresonant case $\Delta\neq0$ and $f_1\neq f_2$ hoping to find more appropriate set of internal parameters for implementing reversible quantum dynamics.

\section{Algebraic optimization for QMB}
To obtain an equidistant frequency comb without degenerate frequencies and well-controlled dynamics, control over the all system parameters is required.
Herein we assume the initial difference between the frequencies of the resonators and atoms located in them.
Below, we focus on the particular  case of an equidistant frequency comb with a single frequency degeneracy in the middle of the comb (similar to works \cite{Moiseev_2017_PRA,Perminov2019spectral}).
The presence of the spectral degeneracy not only simplifies the analysis, but also provides amazing new possibilities for programming the motherboard.

\begin{figure}[t]
\includegraphics[width = 0.45\textwidth]{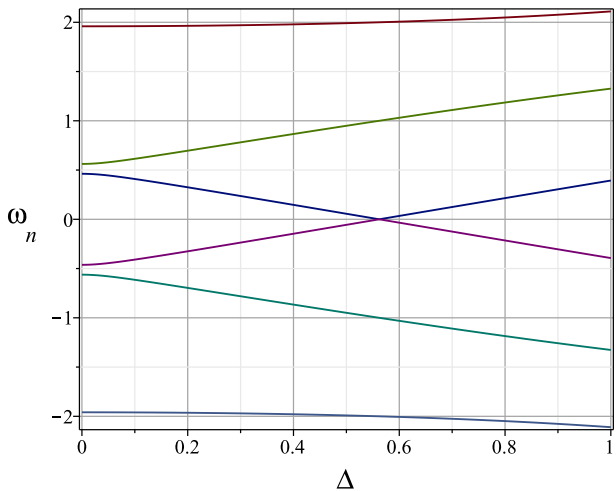}
\caption{Eigenfrequencies versus $\Delta$ for $g=0.7556142107$.}
\label{opt_spectra}
\end{figure}

To obtain the equidistant frequency comb for our QMB, it is necessary to use $\Delta\neq0$ and $f_1\neq f_2$. 
At the beginning, we impose the condition for the presence of degeneracy, which is determined from (\ref{sol_eq}) as the equality $\Delta=f_2$ (see the structure of the factors of the last term in the expression for the $Det$).
Next, we impose the condition for the presence of an equidistant frequency comb in increments of 1, i.e., $\omega_n=\{-2,-1,0,1,2\}$, which definitely gives us the determinant $Det$ structure in the form:
\begin{align}\label{id_det}
Det_0=p^2(p^2+1^2)(p^2+2^2).
\end{align}
Equating the terms with equal powers of $p$ in the expressions for the determinant from (\ref{sol_eq}) and (\ref{id_det}) (condition $Det(p)=Det_0(p)$ for all $p$), we obtain a system of algebraic equations for the parameters $g,\Delta,f_1,f_2$, the solution of which is given by the following optimal parameters:
\begin{align}\label{opt_par}
& \nonumber \Delta=f_2, \\
& \nonumber f_1=2^{-\frac{1}{2}}\sqrt{5-3g^2-\sqrt{g^4-10g^2+9}}, \\
& f_2=2^{-\frac{1}{2}}\sqrt{5-g^2-\sqrt{g^4-10g^2+9}}. 
\end{align}
For such a set of parameters, a complete reversal of the time dynamics will occur at time moment $t=2\pi$. Then after time $t=\pi$ at $g\cong0.7556142107$ there will be a complete transfer of energy from the 1st atom to atoms 2 and 3.
For this interesting case ($g\cong0.7556142107$ and the conditions for $f_1,f_2$ (\ref{opt_par}) are satisfied), the spectrum of the system versus the parameter $\Delta$ is shown in Fig. \ref{opt_spectra}.

We see that the fulfillment of the condition for the presence of an equidistant comb with degeneration is observed only at one point $\Delta\cong0.56206631$. At the same time, it is impossible to achieve the fulfillment of the condition for the presence of an equidistant comb without degeneracy in such a configuration of parameters by controlling only one parameter $\Delta$.
In general, algebraic type solution $s_2(p)$ (\ref{sol_eq}) has very complex spectral properties \cite{Swanson2007,Koziel2011,Tamiazzo2017} and many mathematical aspects of our topic intersect with the fundamental problems of the theory of sensors \cite{Cheng2014} and filters \cite{Swanson2007,Rosenberg2013}.
In particular, conditions such as spectrum degeneracy (the so-called exceptional/diabolic points), that we used earlier to obtain highly efficient quantum memory \cite{Perminov2019spectral}, can have a wide application in the area of photonics \cite{Ozdemir2019parity}.

For our QMB, the condition for determining the degree of root degeneracy and the type of degeneration can be determined analytically based on the Shakirov–-Vieta theorem using the higher discriminants \cite{Shakirov2007}.
Thus, the powerful apparatus of nonlinear algebra \cite{Dolotin2007,Amari2006} allows us to control most of the spectroscopic and dynamic properties of our system on the basis of rapidly computable algebraic feedback. 
This method gives a significant advantage for the rapidly reconfigurable quantum computing platform, which requires a highly accurate method for physically adjusting internal parameters according to the observed spectroscopic data.
We also note that the possibility of algebraic control for ultra-high-dimensional multiparticle quantum systems was opened along this path \cite{Perminov2009discriminants,Morozov2010new,Karasoulou2017algebraic}, which allows creating the distributed scheme of quantum computer in the future with truly great power.

\section{Programmable quantum dynamics}
A unique observation for our platform is the fact that the presence of a comb with frequency degeneration in the central zone corresponding to (\ref{id_det}) does not fix one degree of freedom $g$, while maintaining the relationship between $f_n$ and $g$ (see \ref{opt_par}). This allows programming the energy dynamics of a QMB according to the formula for energy in 2-nd atom $E(x_2)$:
\begin{align}\label{divide_energy}
&\nonumber E_{t=\pi}(x_2)=\frac{(g^4-2g^2+(1-g^2)\sqrt{g^4-10g^2+9}))^2}{9},\\
&\nonumber g(E(x_2)=0)\cong0.7556142107,\\
& g(E(x_2)=1/3)\cong0.4531870484,
\end{align}
which can be used to generate spatially separated logical qubit ($E(x_2)=0,E(x_1)=E(x_3)=1/2$) and qutrit ($E(x_1)=E(x_2)=E(x_3)=1/3$) in one compact platform. When the coupling constant $g$ is quickly disconnected at the time $t=\pi$, these logical qubits can be fixed for further use.
Fast dynamic control of parameter $g$ in such systems is possible due to Josephson junctions and is used to generate entangled states and quantum storage \cite{Flurin2015,Reagor2016quantum}.
For the important case of the complete transfer of energy from the 2nd atom to atoms 1 and 3 (case $E(x_2)=0$ in the formula (\ref{divide_energy})) quantum dynamics near $t=\pi$ is characterized by the broad plateau (an almost rectangular shape).
Due to this temporal behavior, we can perform efficient noise-free operations in subsystems where there is no energy, and we can also perform efficient operations to transform the generated logical qubits.
If we do not dynamically change the parameters of the platform over a period of time $[0;2\pi]$, we get the quantum state storage regime: $E_{t=0}(x_2)=1 \rightarrow E_{t=\pi}(x_2)=0 \rightarrow E_{t=2\pi}(x_2)=1$.
\begin{figure}[t]
\includegraphics[width = 0.45\textwidth]{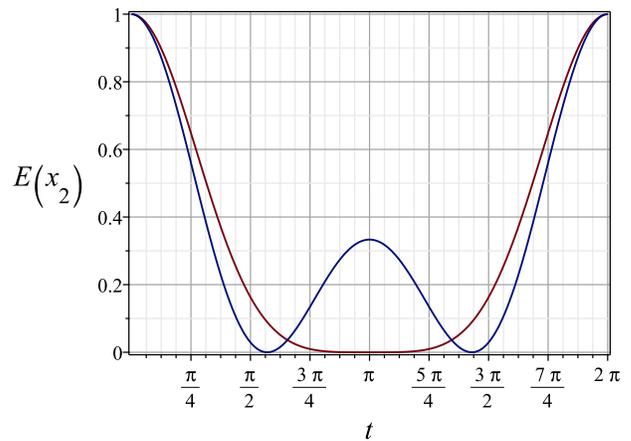}
\caption{Programmable transfer of energy $E(x_2)=|x_2(t)|^2$ in QMB: red line -- $E_{t=\pi}(x_2)=0$, blue line -- $E_{t=\pi}(x_2)=1/3$.}
\label{energy_transfer}
\end{figure}

It can be emphasized that such an effect for the coupled many-particle quantum system is possible only due to the presence of exceptional points determined by zeros by the higher discriminants \cite{Shakirov2007} and 
without special points, programming a highly efficient reversible transfer (while maintaining the structure of the spectrum) is impossible.

\section{Conclusion}
In this paper, we show that multiresonator quantum motherboard scheme allows to achieve an efficient and programmable quantum state transfer between the spatially distributed  subsystems of the multiparticle quantum system.
The controlled transfer is possible for the quantum system with equidistant eigenfreqencies created at special choice of the coupling constants and initial atomic and resonator frequencies.
The studied scheme can be implemented on the studied systems of high-quality whispering gallery modes microresonators \cite{Gorodetsky1994,Gorodetsky1999} coupled to the quantum dots, NV-centers in diamond \cite{Jiang2009} and rare-earth ions \cite{Zhong2015}. 
Herein, the coherent control of the optical atomic coherence can be carried out by an additional lasers tuned to other eigenfrequencies of microresonators.

The predicted dynamics extends the algorithmic capabilities of quantum computation with logical qubits and qutrits. The optimization of all the parameters in QMB with 3 resonators is possible for atomic parameters covering a wide frequency band which provides fast operation rate.
We note that the optimization of the multiparticle dynamics can be performed analytically in more general way on the basis of applied methods of algebraic geometry \cite{Amari2006,Shakirov2007,Dolotin2007}, in particular, for the application of our QMB scheme in large-scale quantum computation  with many spatial channels.
The proposed QMB can be used to combine several quantum devices into a single broadband multi-qubit pre-processing block with highly controlled dynamic properties \cite{Perminov2019spectral}.
In this way, it is possible to create a scalable hybrid QMB for universal quantum computer with quantum supremacy on the basis of already existing technologies \cite{Brecht_2016,Pierre2014,Du2016,Melloni2010,Romanenko2014,Huet2016,Gu2017}.

\section{Acknowledgments}
Research of quantum interface and quantum memory in the area of photonics and quantum technologies is financially supported by a grant of the Government of the Russian Federation, project No. 14.Z50.31.0040, February 17, 2017 (basic idea, theoretical analysis of quantum dynamics and analysis of results -- S.A.M., N.S.P.). The work is also partially financially supported in the framework of the budgetary theme of the Laboratory of Quantum Optics and Informatics of Zavoisky Physical-Technical Institute within the state assignment No.~0217-2018-0005 (algebraic optimization methods for quantum systems –- N.S.P., D.Y.T.).

\bibliographystyle{apsrev4-1}

\begin{thebibliography}{62}%
\makeatletter
\providecommand \@ifxundefined [1]{%
 \@ifx{#1\undefined}
}%
\providecommand \@ifnum [1]{%
 \ifnum #1\expandafter \@firstoftwo
 \else \expandafter \@secondoftwo
 \fi
}%
\providecommand \@ifx [1]{%
 \ifx #1\expandafter \@firstoftwo
 \else \expandafter \@secondoftwo
 \fi
}%
\providecommand \natexlab [1]{#1}%
\providecommand \enquote  [1]{``#1''}%
\providecommand \bibnamefont  [1]{#1}%
\providecommand \bibfnamefont [1]{#1}%
\providecommand \citenamefont [1]{#1}%
\providecommand \href@noop [0]{\@secondoftwo}%
\providecommand \href [0]{\begingroup \@sanitize@url \@href}%
\providecommand \@href[1]{\@@startlink{#1}\@@href}%
\providecommand \@@href[1]{\endgroup#1\@@endlink}%
\providecommand \@sanitize@url [0]{\catcode `\\12\catcode `\$12\catcode
  `\&12\catcode `\#12\catcode `\^12\catcode `\_12\catcode `\%12\relax}%
\providecommand \@@startlink[1]{}%
\providecommand \@@endlink[0]{}%
\providecommand \url  [0]{\begingroup\@sanitize@url \@url }%
\providecommand \@url [1]{\endgroup\@href {#1}{\urlprefix }}%
\providecommand \urlprefix  [0]{URL }%
\providecommand \Eprint [0]{\href }%
\providecommand \doibase [0]{http://dx.doi.org/}%
\providecommand \selectlanguage [0]{\@gobble}%
\providecommand \bibinfo  [0]{\@secondoftwo}%
\providecommand \bibfield  [0]{\@secondoftwo}%
\providecommand \translation [1]{[#1]}%
\providecommand \BibitemOpen [0]{}%
\providecommand \bibitemStop [0]{}%
\providecommand \bibitemNoStop [0]{.\EOS\space}%
\providecommand \EOS [0]{\spacefactor3000\relax}%
\providecommand \BibitemShut  [1]{\csname bibitem#1\endcsname}%
\let\auto@bib@innerbib\@empty
\bibitem [{\citenamefont {Hammerer}\ \emph {et~al.}(2010)\citenamefont
  {Hammerer}, \citenamefont {S\o{}rensen},\ and\ \citenamefont
  {Polzik}}]{Hammerer2010}%
  \BibitemOpen
  \bibfield  {author} {\bibinfo {author} {\bibfnamefont {K.}~\bibnamefont
  {Hammerer}}, \bibinfo {author} {\bibfnamefont {A.~S.}\ \bibnamefont
  {S\o{}rensen}}, \ and\ \bibinfo {author} {\bibfnamefont {E.~S.}\ \bibnamefont
  {Polzik}},\ }\href {\doibase 10.1103/RevModPhys.82.1041} {\bibfield
  {journal} {\bibinfo  {journal} {Rev. Mod. Phys.}\ }\textbf {\bibinfo {volume}
  {82}},\ \bibinfo {pages} {1041} (\bibinfo {year} {2010})}\BibitemShut
  {NoStop}%
\bibitem [{\citenamefont {Kurizki}\ \emph {et~al.}(2015)\citenamefont
  {Kurizki}, \citenamefont {Bertet}, \citenamefont {Kubo}, \citenamefont
  {M\o{}lmer}, \citenamefont {Petrosyan}, \citenamefont {Rabl},\ and\
  \citenamefont {Schmiedmayer}}]{Kurizki2015}%
  \BibitemOpen
  \bibfield  {author} {\bibinfo {author} {\bibfnamefont {G.}~\bibnamefont
  {Kurizki}}, \bibinfo {author} {\bibfnamefont {P.}~\bibnamefont {Bertet}},
  \bibinfo {author} {\bibfnamefont {Y.}~\bibnamefont {Kubo}}, \bibinfo {author}
  {\bibfnamefont {K.}~\bibnamefont {M\o{}lmer}}, \bibinfo {author}
  {\bibfnamefont {D.}~\bibnamefont {Petrosyan}}, \bibinfo {author}
  {\bibfnamefont {P.}~\bibnamefont {Rabl}}, \ and\ \bibinfo {author}
  {\bibfnamefont {J.}~\bibnamefont {Schmiedmayer}},\ }\href@noop {} {\bibfield
  {journal} {\bibinfo  {journal} {Proceedings of the National Academy of
  Sciences}\ }\textbf {\bibinfo {volume} {112}},\ \bibinfo {pages} {3866}
  (\bibinfo {year} {2015})}\BibitemShut {NoStop}%
\bibitem [{\citenamefont {Roy}\ \emph {et~al.}(2017)\citenamefont {Roy},
  \citenamefont {Wilson},\ and\ \citenamefont {Firstenberg}}]{Roy2017}%
  \BibitemOpen
  \bibfield  {author} {\bibinfo {author} {\bibfnamefont {D.}~\bibnamefont
  {Roy}}, \bibinfo {author} {\bibfnamefont {C.~M.}\ \bibnamefont {Wilson}}, \
  and\ \bibinfo {author} {\bibfnamefont {O.}~\bibnamefont {Firstenberg}},\
  }\href {\doibase 10.1103/RevModPhys.89.021001} {\bibfield  {journal}
  {\bibinfo  {journal} {Rev. Mod. Phys.}\ }\textbf {\bibinfo {volume} {89}},\
  \bibinfo {pages} {021001} (\bibinfo {year} {2017})}\BibitemShut {NoStop}%
\bibitem [{\citenamefont {Jiang}\ \emph {et~al.}(2009)\citenamefont {Jiang},
  \citenamefont {Hodges}, \citenamefont {Maze}, \citenamefont {Maurer},
  \citenamefont {Taylor}, \citenamefont {Cory}, \citenamefont {Hemmer},
  \citenamefont {Walsworth}, \citenamefont {Yacoby}, \citenamefont {Zibrov},\
  and\ \citenamefont {Lukin}}]{Jiang2009}%
  \BibitemOpen
  \bibfield  {author} {\bibinfo {author} {\bibfnamefont {L.}~\bibnamefont
  {Jiang}}, \bibinfo {author} {\bibfnamefont {J.~S.}\ \bibnamefont {Hodges}},
  \bibinfo {author} {\bibfnamefont {J.~R.}\ \bibnamefont {Maze}}, \bibinfo
  {author} {\bibfnamefont {P.}~\bibnamefont {Maurer}}, \bibinfo {author}
  {\bibfnamefont {J.~M.}\ \bibnamefont {Taylor}}, \bibinfo {author}
  {\bibfnamefont {D.~G.}\ \bibnamefont {Cory}}, \bibinfo {author}
  {\bibfnamefont {P.~R.}\ \bibnamefont {Hemmer}}, \bibinfo {author}
  {\bibfnamefont {R.~L.}\ \bibnamefont {Walsworth}}, \bibinfo {author}
  {\bibfnamefont {A.}~\bibnamefont {Yacoby}}, \bibinfo {author} {\bibfnamefont
  {A.~S.}\ \bibnamefont {Zibrov}}, \ and\ \bibinfo {author} {\bibfnamefont
  {M.~D.}\ \bibnamefont {Lukin}},\ }\href {\doibase 10.1126/science.1176496}
  {\bibfield  {journal} {\bibinfo  {journal} {Science}\ }\textbf {\bibinfo
  {volume} {326}},\ \bibinfo {pages} {267} (\bibinfo {year}
  {2009})}\BibitemShut {NoStop}%
\bibitem [{\citenamefont {Zhong}\ \emph {et~al.}(2015)\citenamefont {Zhong},
  \citenamefont {Hedges}, \citenamefont {Ahlefeldt}, \citenamefont
  {Bartholomew}, \citenamefont {Beavan}, \citenamefont {Wittig}, \citenamefont
  {Longdell},\ and\ \citenamefont {Sellars}}]{Zhong2015}%
  \BibitemOpen
  \bibfield  {author} {\bibinfo {author} {\bibfnamefont {M.}~\bibnamefont
  {Zhong}}, \bibinfo {author} {\bibfnamefont {M.~P.}\ \bibnamefont {Hedges}},
  \bibinfo {author} {\bibfnamefont {R.~L.}\ \bibnamefont {Ahlefeldt}}, \bibinfo
  {author} {\bibfnamefont {J.~G.}\ \bibnamefont {Bartholomew}}, \bibinfo
  {author} {\bibfnamefont {S.~E.}\ \bibnamefont {Beavan}}, \bibinfo {author}
  {\bibfnamefont {S.~M.}\ \bibnamefont {Wittig}}, \bibinfo {author}
  {\bibfnamefont {J.~J.}\ \bibnamefont {Longdell}}, \ and\ \bibinfo {author}
  {\bibfnamefont {M.~J.}\ \bibnamefont {Sellars}},\ }\href {\doibase
  10.1038/nature14025} {\bibfield  {journal} {\bibinfo  {journal} {Nature}\
  }\textbf {\bibinfo {volume} {517}},\ \bibinfo {pages} {177} (\bibinfo {year}
  {2015})}\BibitemShut {NoStop}%
\bibitem [{\citenamefont {Zhang}\ \emph {et~al.}(2018)\citenamefont {Zhang},
  \citenamefont {Li}, \citenamefont {Cao}, \citenamefont {Xiao}, \citenamefont
  {Guo},\ and\ \citenamefont {Guo}}]{Zhang2019}%
  \BibitemOpen
  \bibfield  {author} {\bibinfo {author} {\bibfnamefont {X.}~\bibnamefont
  {Zhang}}, \bibinfo {author} {\bibfnamefont {H.-O.}\ \bibnamefont {Li}},
  \bibinfo {author} {\bibfnamefont {G.}~\bibnamefont {Cao}}, \bibinfo {author}
  {\bibfnamefont {M.}~\bibnamefont {Xiao}}, \bibinfo {author} {\bibfnamefont
  {G.-C.}\ \bibnamefont {Guo}}, \ and\ \bibinfo {author} {\bibfnamefont
  {G.-P.}\ \bibnamefont {Guo}},\ }\href {\doibase 10.1093/nsr/nwy153}
  {\bibfield  {journal} {\bibinfo  {journal} {National Science Review}\
  }\textbf {\bibinfo {volume} {6}},\ \bibinfo {pages} {32} (\bibinfo {year}
  {2018})}\BibitemShut {NoStop}%
\bibitem [{\citenamefont {Cho}\ \emph {et~al.}(2016)\citenamefont {Cho},
  \citenamefont {Campbell}, \citenamefont {Everett}, \citenamefont {Bernu},
  \citenamefont {Higginbottom}, \citenamefont {Cao}, \citenamefont {Geng},
  \citenamefont {Robins}, \citenamefont {Lam},\ and\ \citenamefont
  {Buchler}}]{Cho2016}%
  \BibitemOpen
  \bibfield  {author} {\bibinfo {author} {\bibfnamefont {Y.-W.}\ \bibnamefont
  {Cho}}, \bibinfo {author} {\bibfnamefont {G.~T.}\ \bibnamefont {Campbell}},
  \bibinfo {author} {\bibfnamefont {J.~L.}\ \bibnamefont {Everett}}, \bibinfo
  {author} {\bibfnamefont {J.}~\bibnamefont {Bernu}}, \bibinfo {author}
  {\bibfnamefont {D.~B.}\ \bibnamefont {Higginbottom}}, \bibinfo {author}
  {\bibfnamefont {M.~T.}\ \bibnamefont {Cao}}, \bibinfo {author} {\bibfnamefont
  {J.}~\bibnamefont {Geng}}, \bibinfo {author} {\bibfnamefont {N.~P.}\
  \bibnamefont {Robins}}, \bibinfo {author} {\bibfnamefont {P.~K.}\
  \bibnamefont {Lam}}, \ and\ \bibinfo {author} {\bibfnamefont {B.~C.}\
  \bibnamefont {Buchler}},\ }\href {\doibase 10.1364/OPTICA.3.000100}
  {\bibfield  {journal} {\bibinfo  {journal} {Optica}\ }\textbf {\bibinfo
  {volume} {3}},\ \bibinfo {pages} {100} (\bibinfo {year} {2016})}\BibitemShut
  {NoStop}%
\bibitem [{\citenamefont {Hsiao}\ \emph {et~al.}(2018)\citenamefont {Hsiao},
  \citenamefont {Tsai}, \citenamefont {Chen}, \citenamefont {Lin},
  \citenamefont {Hung}, \citenamefont {Lee}, \citenamefont {Chen},
  \citenamefont {Chen}, \citenamefont {Yu},\ and\ \citenamefont
  {Chen}}]{Hsiao2018}%
  \BibitemOpen
  \bibfield  {author} {\bibinfo {author} {\bibfnamefont {Y.-F.}\ \bibnamefont
  {Hsiao}}, \bibinfo {author} {\bibfnamefont {P.-J.}\ \bibnamefont {Tsai}},
  \bibinfo {author} {\bibfnamefont {H.-S.}\ \bibnamefont {Chen}}, \bibinfo
  {author} {\bibfnamefont {S.-X.}\ \bibnamefont {Lin}}, \bibinfo {author}
  {\bibfnamefont {C.-C.}\ \bibnamefont {Hung}}, \bibinfo {author}
  {\bibfnamefont {C.-H.}\ \bibnamefont {Lee}}, \bibinfo {author} {\bibfnamefont
  {Y.-H.}\ \bibnamefont {Chen}}, \bibinfo {author} {\bibfnamefont {Y.-F.}\
  \bibnamefont {Chen}}, \bibinfo {author} {\bibfnamefont {I.~A.}\ \bibnamefont
  {Yu}}, \ and\ \bibinfo {author} {\bibfnamefont {Y.-C.}\ \bibnamefont
  {Chen}},\ }\href {\doibase 10.1103/PhysRevLett.120.183602} {\bibfield
  {journal} {\bibinfo  {journal} {Phys. Rev. Lett.}\ }\textbf {\bibinfo
  {volume} {120}},\ \bibinfo {pages} {183602} (\bibinfo {year}
  {2018})}\BibitemShut {NoStop}%
\bibitem [{\citenamefont {Hartmann}\ \emph {et~al.}(2008)\citenamefont
  {Hartmann}, \citenamefont {BrandГЈo},\ and\ \citenamefont
  {Plenio}}]{Hartmann2008}%
  \BibitemOpen
  \bibfield  {author} {\bibinfo {author} {\bibfnamefont {M.}~\bibnamefont
  {Hartmann}}, \bibinfo {author} {\bibfnamefont {F.}~\bibnamefont {BrandГЈo}},
  \ and\ \bibinfo {author} {\bibfnamefont {M.}~\bibnamefont {Plenio}},\ }\href
  {\doibase 10.1002/lpor.200810046} {\bibfield  {journal} {\bibinfo  {journal}
  {Laser {\&} Photonics Reviews}\ }\textbf {\bibinfo {volume} {2}},\ \bibinfo
  {pages} {527} (\bibinfo {year} {2008})}\BibitemShut {NoStop}%
\bibitem [{\citenamefont {Hur}\ \emph {et~al.}(2016)\citenamefont {Hur},
  \citenamefont {Henriet}, \citenamefont {Petrescu}, \citenamefont {Plekhanov},
  \citenamefont {Roux},\ and\ \citenamefont {SchirГі}}]{Hur2016}%
  \BibitemOpen
  \bibfield  {author} {\bibinfo {author} {\bibfnamefont {K.~L.}\ \bibnamefont
  {Hur}}, \bibinfo {author} {\bibfnamefont {L.}~\bibnamefont {Henriet}},
  \bibinfo {author} {\bibfnamefont {A.}~\bibnamefont {Petrescu}}, \bibinfo
  {author} {\bibfnamefont {K.}~\bibnamefont {Plekhanov}}, \bibinfo {author}
  {\bibfnamefont {G.}~\bibnamefont {Roux}}, \ and\ \bibinfo {author}
  {\bibfnamefont {M.}~\bibnamefont {SchirГі}},\ }\href {\doibase
  10.1016/j.crhy.2016.05.003} {\bibfield  {journal} {\bibinfo  {journal}
  {Comptes Rendus Physique}\ }\textbf {\bibinfo {volume} {17}},\ \bibinfo
  {pages} {808 } (\bibinfo {year} {2016})}\BibitemShut {NoStop}%
\bibitem [{\citenamefont {Noh}\ and\ \citenamefont
  {Angelakis}(2017)}]{Noh2017}%
  \BibitemOpen
  \bibfield  {author} {\bibinfo {author} {\bibfnamefont {C.}~\bibnamefont
  {Noh}}\ and\ \bibinfo {author} {\bibfnamefont {D.~G.}\ \bibnamefont
  {Angelakis}},\ }\href {\doibase 10.1088/0034-4885/80/1/016401} {\bibfield
  {journal} {\bibinfo  {journal} {Reports on Progress in Physics}\ }\textbf
  {\bibinfo {volume} {80}},\ \bibinfo {pages} {016401} (\bibinfo {year}
  {2017})}\BibitemShut {NoStop}%
\bibitem [{\citenamefont {Mirhosseini}\ \emph {et~al.}(2019)\citenamefont
  {Mirhosseini}, \citenamefont {Kim}, \citenamefont {Zhang}, \citenamefont
  {Sipahigil}, \citenamefont {Dieterle}, \citenamefont {Keller}, \citenamefont
  {Asenjo-Garcia}, \citenamefont {Chang},\ and\ \citenamefont
  {Painter}}]{Mirhosseini2019cavity}%
  \BibitemOpen
  \bibfield  {author} {\bibinfo {author} {\bibfnamefont {M.}~\bibnamefont
  {Mirhosseini}}, \bibinfo {author} {\bibfnamefont {E.}~\bibnamefont {Kim}},
  \bibinfo {author} {\bibfnamefont {X.}~\bibnamefont {Zhang}}, \bibinfo
  {author} {\bibfnamefont {A.}~\bibnamefont {Sipahigil}}, \bibinfo {author}
  {\bibfnamefont {P.~B.}\ \bibnamefont {Dieterle}}, \bibinfo {author}
  {\bibfnamefont {A.~J.}\ \bibnamefont {Keller}}, \bibinfo {author}
  {\bibfnamefont {A.}~\bibnamefont {Asenjo-Garcia}}, \bibinfo {author}
  {\bibfnamefont {D.~E.}\ \bibnamefont {Chang}}, \ and\ \bibinfo {author}
  {\bibfnamefont {O.}~\bibnamefont {Painter}},\ }\href {\doibase
  10.1038/s41586-019-1196-1} {\bibfield  {journal} {\bibinfo  {journal}
  {Nature}\ }\textbf {\bibinfo {volume} {569}},\ \bibinfo {pages} {692}
  (\bibinfo {year} {2019})}\BibitemShut {NoStop}%
\bibitem [{\citenamefont {Gorodetsky}\ and\ \citenamefont
  {Ilchenko}(1999)}]{Gorodetsky1999}%
  \BibitemOpen
  \bibfield  {author} {\bibinfo {author} {\bibfnamefont {M.~L.}\ \bibnamefont
  {Gorodetsky}}\ and\ \bibinfo {author} {\bibfnamefont {V.~S.}\ \bibnamefont
  {Ilchenko}},\ }\href {\doibase 10.1364/JOSAB.16.000147} {\bibfield  {journal}
  {\bibinfo  {journal} {J. Opt. Soc. Am. B}\ }\textbf {\bibinfo {volume}
  {16}},\ \bibinfo {pages} {147} (\bibinfo {year} {1999})}\BibitemShut
  {NoStop}%
\bibitem [{\citenamefont {Vahala}(2003)}]{Vahala2003}%
  \BibitemOpen
  \bibfield  {author} {\bibinfo {author} {\bibfnamefont {K.~J.}\ \bibnamefont
  {Vahala}},\ }\href {\doibase 10.1038/nature01939} {\bibfield  {journal}
  {\bibinfo  {journal} {Nature}\ }\textbf {\bibinfo {volume} {424}},\ \bibinfo
  {pages} {839 EP} (\bibinfo {year} {2003})}\BibitemShut {NoStop}%
\bibitem [{\citenamefont {Kobe}\ \emph {et~al.}(2017)\citenamefont {Kobe},
  \citenamefont {Chuma}, \citenamefont {Jr.},\ and\ \citenamefont
  {Chose}}]{Kobe2017}%
  \BibitemOpen
  \bibfield  {author} {\bibinfo {author} {\bibfnamefont {O.~B.}\ \bibnamefont
  {Kobe}}, \bibinfo {author} {\bibfnamefont {J.}~\bibnamefont {Chuma}},
  \bibinfo {author} {\bibfnamefont {R.~J.}\ \bibnamefont {Jr.}}, \ and\
  \bibinfo {author} {\bibfnamefont {M.}~\bibnamefont {Chose}},\ }\href
  {\doibase 10.1016/j.jestch.2016.09.024} {\bibfield  {journal} {\bibinfo
  {journal} {Engineering Science and Technology}\ }\textbf {\bibinfo {volume}
  {20}},\ \bibinfo {pages} {460 } (\bibinfo {year} {2017})}\BibitemShut
  {NoStop}%
\bibitem [{\citenamefont {Toth}\ \emph {et~al.}(2017)\citenamefont {Toth},
  \citenamefont {Bernier}, \citenamefont {Nunnenkamp}, \citenamefont
  {Feofanov},\ and\ \citenamefont {Kippenberg}}]{Toth2017}%
  \BibitemOpen
  \bibfield  {author} {\bibinfo {author} {\bibfnamefont {L.~D.}\ \bibnamefont
  {Toth}}, \bibinfo {author} {\bibfnamefont {N.~R.}\ \bibnamefont {Bernier}},
  \bibinfo {author} {\bibfnamefont {A.}~\bibnamefont {Nunnenkamp}}, \bibinfo
  {author} {\bibfnamefont {A.~K.}\ \bibnamefont {Feofanov}}, \ and\ \bibinfo
  {author} {\bibfnamefont {T.~J.}\ \bibnamefont {Kippenberg}},\ }\href
  {\doibase 10.1038/nphys4121} {\bibfield  {journal} {\bibinfo  {journal}
  {Nature Physics}\ }\textbf {\bibinfo {volume} {13}},\ \bibinfo {pages} {787}
  (\bibinfo {year} {2017})}\BibitemShut {NoStop}%
\bibitem [{\citenamefont {Megrant}\ \emph {et~al.}(2012)\citenamefont
  {Megrant}, \citenamefont {Neill}, \citenamefont {Barends}, \citenamefont
  {Chiaro}, \citenamefont {Chen}, \citenamefont {Feigl}, \citenamefont {Kelly},
  \citenamefont {Lucero}, \citenamefont {Mariantoni}, \citenamefont
  {OвЂ™Malley}, \citenamefont {Sank}, \citenamefont {Vainsencher},
  \citenamefont {Wenner}, \citenamefont {White}, \citenamefont {Yin},
  \citenamefont {Zhao}, \citenamefont {Palmstr{\"o}m}, \citenamefont
  {Martinis},\ and\ \citenamefont {Cleland}}]{Megrant2012}%
  \BibitemOpen
  \bibfield  {author} {\bibinfo {author} {\bibfnamefont {A.}~\bibnamefont
  {Megrant}}, \bibinfo {author} {\bibfnamefont {C.}~\bibnamefont {Neill}},
  \bibinfo {author} {\bibfnamefont {R.}~\bibnamefont {Barends}}, \bibinfo
  {author} {\bibfnamefont {B.}~\bibnamefont {Chiaro}}, \bibinfo {author}
  {\bibfnamefont {Y.}~\bibnamefont {Chen}}, \bibinfo {author} {\bibfnamefont
  {L.}~\bibnamefont {Feigl}}, \bibinfo {author} {\bibfnamefont
  {J.}~\bibnamefont {Kelly}}, \bibinfo {author} {\bibfnamefont
  {E.}~\bibnamefont {Lucero}}, \bibinfo {author} {\bibfnamefont
  {M.}~\bibnamefont {Mariantoni}}, \bibinfo {author} {\bibfnamefont {P.~J.~J.}\
  \bibnamefont {OвЂ™Malley}}, \bibinfo {author} {\bibfnamefont
  {D.}~\bibnamefont {Sank}}, \bibinfo {author} {\bibfnamefont {A.}~\bibnamefont
  {Vainsencher}}, \bibinfo {author} {\bibfnamefont {J.}~\bibnamefont {Wenner}},
  \bibinfo {author} {\bibfnamefont {T.~C.}\ \bibnamefont {White}}, \bibinfo
  {author} {\bibfnamefont {Y.}~\bibnamefont {Yin}}, \bibinfo {author}
  {\bibfnamefont {J.}~\bibnamefont {Zhao}}, \bibinfo {author} {\bibfnamefont
  {C.~J.}\ \bibnamefont {Palmstr{\"o}m}}, \bibinfo {author} {\bibfnamefont
  {J.~M.}\ \bibnamefont {Martinis}}, \ and\ \bibinfo {author} {\bibfnamefont
  {A.~N.}\ \bibnamefont {Cleland}},\ }\href {\doibase 10.1063/1.3693409}
  {\bibfield  {journal} {\bibinfo  {journal} {Applied Physics Letters}\
  }\textbf {\bibinfo {volume} {100}},\ \bibinfo {pages} {113510} (\bibinfo
  {year} {2012})}\BibitemShut {NoStop}%
\bibitem [{\citenamefont {Armani}\ \emph {et~al.}(2003)\citenamefont {Armani},
  \citenamefont {Kippenberg}, \citenamefont {Spillane},\ and\ \citenamefont
  {Vahala}}]{Armani2003}%
  \BibitemOpen
  \bibfield  {author} {\bibinfo {author} {\bibfnamefont {D.~K.}\ \bibnamefont
  {Armani}}, \bibinfo {author} {\bibfnamefont {T.~J.}\ \bibnamefont
  {Kippenberg}}, \bibinfo {author} {\bibfnamefont {S.~M.}\ \bibnamefont
  {Spillane}}, \ and\ \bibinfo {author} {\bibfnamefont {K.~J.}\ \bibnamefont
  {Vahala}},\ }\href {\doibase 10.1038/nature01371} {\bibfield  {journal}
  {\bibinfo  {journal} {Nature}\ }\textbf {\bibinfo {volume} {421}},\ \bibinfo
  {pages} {925 EP} (\bibinfo {year} {2003})}\BibitemShut {NoStop}%
\bibitem [{\citenamefont {Liu}\ \emph {et~al.}(2018)\citenamefont {Liu},
  \citenamefont {Guo}, \citenamefont {Zhang}, \citenamefont {Yu},\ and\
  \citenamefont {Zhang}}]{Liu2018}%
  \BibitemOpen
  \bibfield  {author} {\bibinfo {author} {\bibfnamefont {T.}~\bibnamefont
  {Liu}}, \bibinfo {author} {\bibfnamefont {B.-Q.}\ \bibnamefont {Guo}},
  \bibinfo {author} {\bibfnamefont {Y.}~\bibnamefont {Zhang}}, \bibinfo
  {author} {\bibfnamefont {C.-S.}\ \bibnamefont {Yu}}, \ and\ \bibinfo {author}
  {\bibfnamefont {W.-N.}\ \bibnamefont {Zhang}},\ }\href {\doibase
  10.1007/s11128-018-2011-x} {\bibfield  {journal} {\bibinfo  {journal}
  {Quantum Information Processing}\ }\textbf {\bibinfo {volume} {17}},\
  \bibinfo {pages} {240} (\bibinfo {year} {2018})}\BibitemShut {NoStop}%
\bibitem [{\citenamefont {Xie}\ \emph {et~al.}(2018)\citenamefont {Xie},
  \citenamefont {Deppe}, \citenamefont {Renger}, \citenamefont {Repp},
  \citenamefont {Eder}, \citenamefont {Fischer}, \citenamefont {Goetz},
  \citenamefont {Pogorzalek}, \citenamefont {Fedorov}, \citenamefont {Marx},\
  and\ \citenamefont {Gross}}]{Xie2018}%
  \BibitemOpen
  \bibfield  {author} {\bibinfo {author} {\bibfnamefont {E.}~\bibnamefont
  {Xie}}, \bibinfo {author} {\bibfnamefont {F.}~\bibnamefont {Deppe}}, \bibinfo
  {author} {\bibfnamefont {M.}~\bibnamefont {Renger}}, \bibinfo {author}
  {\bibfnamefont {D.}~\bibnamefont {Repp}}, \bibinfo {author} {\bibfnamefont
  {P.}~\bibnamefont {Eder}}, \bibinfo {author} {\bibfnamefont {M.}~\bibnamefont
  {Fischer}}, \bibinfo {author} {\bibfnamefont {J.}~\bibnamefont {Goetz}},
  \bibinfo {author} {\bibfnamefont {S.}~\bibnamefont {Pogorzalek}}, \bibinfo
  {author} {\bibfnamefont {K.~G.}\ \bibnamefont {Fedorov}}, \bibinfo {author}
  {\bibfnamefont {A.}~\bibnamefont {Marx}}, \ and\ \bibinfo {author}
  {\bibfnamefont {R.}~\bibnamefont {Gross}},\ }\href {\doibase
  10.1063/1.5029514} {\bibfield  {journal} {\bibinfo  {journal} {Applied
  Physics Letters}\ }\textbf {\bibinfo {volume} {112}},\ \bibinfo {pages}
  {202601} (\bibinfo {year} {2018})}\BibitemShut {NoStop}%
\bibitem [{\citenamefont {Flurin}\ \emph {et~al.}(2015)\citenamefont {Flurin},
  \citenamefont {Roch}, \citenamefont {Pillet}, \citenamefont {Mallet},\ and\
  \citenamefont {Huard}}]{Flurin2015}%
  \BibitemOpen
  \bibfield  {author} {\bibinfo {author} {\bibfnamefont {E.}~\bibnamefont
  {Flurin}}, \bibinfo {author} {\bibfnamefont {N.}~\bibnamefont {Roch}},
  \bibinfo {author} {\bibfnamefont {J.~D.}\ \bibnamefont {Pillet}}, \bibinfo
  {author} {\bibfnamefont {F.}~\bibnamefont {Mallet}}, \ and\ \bibinfo {author}
  {\bibfnamefont {B.}~\bibnamefont {Huard}},\ }\href {\doibase
  10.1103/PhysRevLett.114.090503} {\bibfield  {journal} {\bibinfo  {journal}
  {Phys. Rev. Lett.}\ }\textbf {\bibinfo {volume} {114}},\ \bibinfo {pages}
  {090503} (\bibinfo {year} {2015})}\BibitemShut {NoStop}%
\bibitem [{\citenamefont {Pfaff}\ \emph {et~al.}(2017)\citenamefont {Pfaff},
  \citenamefont {Axline}, \citenamefont {Burkhart}, \citenamefont {Vool},
  \citenamefont {Reinhold}, \citenamefont {Frunzio}, \citenamefont {Jiang},
  \citenamefont {Devoret},\ and\ \citenamefont {Schoelkopf}}]{Pfaff2017}%
  \BibitemOpen
  \bibfield  {author} {\bibinfo {author} {\bibfnamefont {W.}~\bibnamefont
  {Pfaff}}, \bibinfo {author} {\bibfnamefont {C.~J.}\ \bibnamefont {Axline}},
  \bibinfo {author} {\bibfnamefont {L.~D.}\ \bibnamefont {Burkhart}}, \bibinfo
  {author} {\bibfnamefont {U.}~\bibnamefont {Vool}}, \bibinfo {author}
  {\bibfnamefont {P.}~\bibnamefont {Reinhold}}, \bibinfo {author}
  {\bibfnamefont {L.}~\bibnamefont {Frunzio}}, \bibinfo {author} {\bibfnamefont
  {L.}~\bibnamefont {Jiang}}, \bibinfo {author} {\bibfnamefont {M.~H.}\
  \bibnamefont {Devoret}}, \ and\ \bibinfo {author} {\bibfnamefont {R.~J.}\
  \bibnamefont {Schoelkopf}},\ }\href@noop {} {\bibfield  {journal} {\bibinfo
  {journal} {Nature Physics}\ } (\bibinfo {year} {2017})}\BibitemShut {NoStop}%
\bibitem [{\citenamefont {Sirois}\ \emph {et~al.}(2015)\citenamefont {Sirois},
  \citenamefont {Castellanos-Beltran}, \citenamefont {DeFeo}, \citenamefont
  {Ranzani}, \citenamefont {Lecocq}, \citenamefont {Simmonds}, \citenamefont
  {Teufel},\ and\ \citenamefont {Aumentado}}]{Sirois2017}%
  \BibitemOpen
  \bibfield  {author} {\bibinfo {author} {\bibfnamefont {A.~J.}\ \bibnamefont
  {Sirois}}, \bibinfo {author} {\bibfnamefont {M.~A.}\ \bibnamefont
  {Castellanos-Beltran}}, \bibinfo {author} {\bibfnamefont {M.~P.}\
  \bibnamefont {DeFeo}}, \bibinfo {author} {\bibfnamefont {L.}~\bibnamefont
  {Ranzani}}, \bibinfo {author} {\bibfnamefont {F.}~\bibnamefont {Lecocq}},
  \bibinfo {author} {\bibfnamefont {R.~W.}\ \bibnamefont {Simmonds}}, \bibinfo
  {author} {\bibfnamefont {J.~D.}\ \bibnamefont {Teufel}}, \ and\ \bibinfo
  {author} {\bibfnamefont {J.}~\bibnamefont {Aumentado}},\ }\href {\doibase
  10.1063/1.4919759} {\bibfield  {journal} {\bibinfo  {journal} {Applied
  Physics Letters}\ }\textbf {\bibinfo {volume} {106}},\ \bibinfo {pages}
  {172603} (\bibinfo {year} {2015})}\BibitemShut {NoStop}%
\bibitem [{\citenamefont {Moiseev}\ and\ \citenamefont
  {Moiseev}(2017)}]{EMoiseev2017}%
  \BibitemOpen
  \bibfield  {author} {\bibinfo {author} {\bibfnamefont {E.~S.}\ \bibnamefont
  {Moiseev}}\ and\ \bibinfo {author} {\bibfnamefont {S.~A.}\ \bibnamefont
  {Moiseev}},\ }\href {\doibase 10.1088/1612-202X/aa4fc2} {\bibfield  {journal}
  {\bibinfo  {journal} {Laser Physics Letters}\ }\textbf {\bibinfo {volume}
  {14}},\ \bibinfo {pages} {015202} (\bibinfo {year} {2017})}\BibitemShut
  {NoStop}%
\bibitem [{\citenamefont {Moiseev}\ \emph {et~al.}(2017)\citenamefont
  {Moiseev}, \citenamefont {Gubaidullin}, \citenamefont {Kirillov},
  \citenamefont {Latypov}, \citenamefont {Perminov}, \citenamefont
  {Petrovnin},\ and\ \citenamefont {Sherstyukov}}]{Moiseev_2017_PRA}%
  \BibitemOpen
  \bibfield  {author} {\bibinfo {author} {\bibfnamefont {S.~A.}\ \bibnamefont
  {Moiseev}}, \bibinfo {author} {\bibfnamefont {F.~F.}\ \bibnamefont
  {Gubaidullin}}, \bibinfo {author} {\bibfnamefont {R.~S.}\ \bibnamefont
  {Kirillov}}, \bibinfo {author} {\bibfnamefont {R.~R.}\ \bibnamefont
  {Latypov}}, \bibinfo {author} {\bibfnamefont {N.~S.}\ \bibnamefont
  {Perminov}}, \bibinfo {author} {\bibfnamefont {K.~V.}\ \bibnamefont
  {Petrovnin}}, \ and\ \bibinfo {author} {\bibfnamefont {O.~N.}\ \bibnamefont
  {Sherstyukov}},\ }\href {\doibase 10.1103/PhysRevA.95.012338} {\bibfield
  {journal} {\bibinfo  {journal} {Physical Review A}\ }\textbf {\bibinfo
  {volume} {95}},\ \bibinfo {pages} {012338} (\bibinfo {year}
  {2017})}\BibitemShut {NoStop}%
\bibitem [{\citenamefont {Moiseev}\ \emph {et~al.}(2018)\citenamefont
  {Moiseev}, \citenamefont {Gerasimov}, \citenamefont {Latypov}, \citenamefont
  {Perminov}, \citenamefont {Petrovnin},\ and\ \citenamefont
  {Sherstyukov}}]{Moiseev2018}%
  \BibitemOpen
  \bibfield  {author} {\bibinfo {author} {\bibfnamefont {S.~A.}\ \bibnamefont
  {Moiseev}}, \bibinfo {author} {\bibfnamefont {K.~I.}\ \bibnamefont
  {Gerasimov}}, \bibinfo {author} {\bibfnamefont {R.~R.}\ \bibnamefont
  {Latypov}}, \bibinfo {author} {\bibfnamefont {N.~S.}\ \bibnamefont
  {Perminov}}, \bibinfo {author} {\bibfnamefont {K.~V.}\ \bibnamefont
  {Petrovnin}}, \ and\ \bibinfo {author} {\bibfnamefont {O.~N.}\ \bibnamefont
  {Sherstyukov}},\ }\href {\doibase 10.1038/s41598-018-21941-6} {\bibfield
  {journal} {\bibinfo  {journal} {Scientific Reports}\ }\textbf {\bibinfo
  {volume} {8}},\ \bibinfo {pages} {3982} (\bibinfo {year} {2018})}\BibitemShut
  {NoStop}%
\bibitem [{\citenamefont {Heebner}\ and\ \citenamefont
  {Boyd}(2002)}]{Heebner2002slow}%
  \BibitemOpen
  \bibfield  {author} {\bibinfo {author} {\bibfnamefont {J.~E.}\ \bibnamefont
  {Heebner}}\ and\ \bibinfo {author} {\bibfnamefont {R.~W.}\ \bibnamefont
  {Boyd}},\ }\href {\doibase 10.1080/0950034021000011527} {\bibfield  {journal}
  {\bibinfo  {journal} {Journal of Modern Optics}\ }\textbf {\bibinfo {volume}
  {49}},\ \bibinfo {pages} {2629} (\bibinfo {year} {2002})}\BibitemShut
  {NoStop}%
\bibitem [{\citenamefont {Perminov}\ \emph {et~al.}(2018)\citenamefont
  {Perminov}, \citenamefont {Tarankova},\ and\ \citenamefont
  {Moiseev}}]{Perminov2018superefficient}%
  \BibitemOpen
  \bibfield  {author} {\bibinfo {author} {\bibfnamefont {N.}~\bibnamefont
  {Perminov}}, \bibinfo {author} {\bibfnamefont {D.~Y.}\ \bibnamefont
  {Tarankova}}, \ and\ \bibinfo {author} {\bibfnamefont {S.}~\bibnamefont
  {Moiseev}},\ }\href {\doibase 10.1088/1612-202X/aae782} {\bibfield  {journal}
  {\bibinfo  {journal} {Laser Physics Letters}\ }\textbf {\bibinfo {volume}
  {15}},\ \bibinfo {pages} {125203} (\bibinfo {year} {2018})}\BibitemShut
  {NoStop}%
\bibitem [{\citenamefont {Perminov}\ and\ \citenamefont
  {Moiseev}(2019)}]{Perminov2019spectral}%
  \BibitemOpen
  \bibfield  {author} {\bibinfo {author} {\bibfnamefont {N.}~\bibnamefont
  {Perminov}}\ and\ \bibinfo {author} {\bibfnamefont {S.}~\bibnamefont
  {Moiseev}},\ }\href {\doibase 10.1038/s41598-018-38244-5} {\bibfield
  {journal} {\bibinfo  {journal} {Scientific reports}\ }\textbf {\bibinfo
  {volume} {9}},\ \bibinfo {pages} {1568} (\bibinfo {year} {2019})}\BibitemShut
  {NoStop}%
\bibitem [{\citenamefont {Kockum}\ \emph {et~al.}(2018)\citenamefont {Kockum},
  \citenamefont {Johansson},\ and\ \citenamefont {Nori}}]{Kockum2018}%
  \BibitemOpen
  \bibfield  {author} {\bibinfo {author} {\bibfnamefont {A.~F.}\ \bibnamefont
  {Kockum}}, \bibinfo {author} {\bibfnamefont {G.}~\bibnamefont {Johansson}}, \
  and\ \bibinfo {author} {\bibfnamefont {F.}~\bibnamefont {Nori}},\ }\href
  {\doibase 10.1103/PhysRevLett.120.140404} {\bibfield  {journal} {\bibinfo
  {journal} {Phys. Rev. Lett.}\ }\textbf {\bibinfo {volume} {120}},\ \bibinfo
  {pages} {140404} (\bibinfo {year} {2018})}\BibitemShut {NoStop}%
\bibitem [{\citenamefont {Moiseev}\ \emph {et~al.}(2013)\citenamefont
  {Moiseev}, \citenamefont {Andrianov},\ and\ \citenamefont
  {Moiseev}}]{Moiseev2013quantum}%
  \BibitemOpen
  \bibfield  {author} {\bibinfo {author} {\bibfnamefont {S.}~\bibnamefont
  {Moiseev}}, \bibinfo {author} {\bibfnamefont {S.}~\bibnamefont {Andrianov}},
  \ and\ \bibinfo {author} {\bibfnamefont {E.}~\bibnamefont {Moiseev}},\ }\href
  {\doibase 10.1134/S0030400X13090166} {\bibfield  {journal} {\bibinfo
  {journal} {Optics and spectroscopy}\ }\textbf {\bibinfo {volume} {115}},\
  \bibinfo {pages} {356} (\bibinfo {year} {2013})}\BibitemShut {NoStop}%
\bibitem [{\citenamefont {Andrianov}\ \emph {et~al.}(2019)\citenamefont
  {Andrianov}, \citenamefont {Arslanov}, \citenamefont {Gerasimov},
  \citenamefont {Kalinkin},\ and\ \citenamefont {Moiseev}}]{Andrianov2019cnot}%
  \BibitemOpen
  \bibfield  {author} {\bibinfo {author} {\bibfnamefont {S.~N.}\ \bibnamefont
  {Andrianov}}, \bibinfo {author} {\bibfnamefont {N.~M.}\ \bibnamefont
  {Arslanov}}, \bibinfo {author} {\bibfnamefont {K.~I.}\ \bibnamefont
  {Gerasimov}}, \bibinfo {author} {\bibfnamefont {A.~A.}\ \bibnamefont
  {Kalinkin}}, \ and\ \bibinfo {author} {\bibfnamefont {S.~A.}\ \bibnamefont
  {Moiseev}},\ }\href {\doibase 10.1007/s11128-019-2345-z} {\bibfield
  {journal} {\bibinfo  {journal} {Quantum Information Processing}\ }\textbf
  {\bibinfo {volume} {18}},\ \bibinfo {pages} {235} (\bibinfo {year}
  {2019})}\BibitemShut {NoStop}%
\bibitem [{\citenamefont {Moiseev}\ and\ \citenamefont
  {Noskov}(2004)}]{Moiseev2004}%
  \BibitemOpen
  \bibfield  {author} {\bibinfo {author} {\bibfnamefont {S.~A.}\ \bibnamefont
  {Moiseev}}\ and\ \bibinfo {author} {\bibfnamefont {M.~I.}\ \bibnamefont
  {Noskov}},\ }\href {\doibase 10.1002/lapl.200310071} {\bibfield  {journal}
  {\bibinfo  {journal} {Laser Physics Letters}\ }\textbf {\bibinfo {volume}
  {1}},\ \bibinfo {pages} {303} (\bibinfo {year} {2004})}\BibitemShut {NoStop}%
\bibitem [{\citenamefont {Tittel}\ \emph {et~al.}(2009)\citenamefont {Tittel},
  \citenamefont {Afzelius}, \citenamefont {Chaneli{\'{e}}re}, \citenamefont
  {Cone}, \citenamefont {Kr{\"{o}}ll}, \citenamefont {Moiseev},\ and\
  \citenamefont {Sellars}}]{Tittel2009}%
  \BibitemOpen
  \bibfield  {author} {\bibinfo {author} {\bibfnamefont {W.}~\bibnamefont
  {Tittel}}, \bibinfo {author} {\bibfnamefont {M.}~\bibnamefont {Afzelius}},
  \bibinfo {author} {\bibfnamefont {T.}~\bibnamefont {Chaneli{\'{e}}re}},
  \bibinfo {author} {\bibfnamefont {R.}~\bibnamefont {Cone}}, \bibinfo {author}
  {\bibfnamefont {S.}~\bibnamefont {Kr{\"{o}}ll}}, \bibinfo {author}
  {\bibfnamefont {S.}~\bibnamefont {Moiseev}}, \ and\ \bibinfo {author}
  {\bibfnamefont {M.}~\bibnamefont {Sellars}},\ }\href {\doibase
  10.1002/lpor.200810056} {\bibfield  {journal} {\bibinfo  {journal} {Laser
  {\&} Photonics Reviews}\ }\textbf {\bibinfo {volume} {4}},\ \bibinfo {pages}
  {244} (\bibinfo {year} {2009})}\BibitemShut {NoStop}%
\bibitem [{\citenamefont {de~Riedmatten}\ \emph {et~al.}(2008)\citenamefont
  {de~Riedmatten}, \citenamefont {Afzelius}, \citenamefont {Staudt},
  \citenamefont {Simon},\ and\ \citenamefont {Gisin}}]{Riedmatten2008}%
  \BibitemOpen
  \bibfield  {author} {\bibinfo {author} {\bibfnamefont {H.}~\bibnamefont
  {de~Riedmatten}}, \bibinfo {author} {\bibfnamefont {M.}~\bibnamefont
  {Afzelius}}, \bibinfo {author} {\bibfnamefont {M.~U.}\ \bibnamefont
  {Staudt}}, \bibinfo {author} {\bibfnamefont {C.}~\bibnamefont {Simon}}, \
  and\ \bibinfo {author} {\bibfnamefont {N.}~\bibnamefont {Gisin}},\ }\href
  {\doibase 10.1038/nature07607} {\bibfield  {journal} {\bibinfo  {journal}
  {Nature}\ }\textbf {\bibinfo {volume} {456}},\ \bibinfo {pages} {773}
  (\bibinfo {year} {2008})}\BibitemShut {NoStop}%
\bibitem [{\citenamefont {Akhmedzhanov}\ \emph {et~al.}(2016)\citenamefont
  {Akhmedzhanov}, \citenamefont {Gushchin}, \citenamefont {Kalachev},
  \citenamefont {Korableva}, \citenamefont {Sobgayda},\ and\ \citenamefont
  {Zelensky}}]{Akhmedzhanov2016}%
  \BibitemOpen
  \bibfield  {author} {\bibinfo {author} {\bibfnamefont {R.~A.}\ \bibnamefont
  {Akhmedzhanov}}, \bibinfo {author} {\bibfnamefont {L.~A.}\ \bibnamefont
  {Gushchin}}, \bibinfo {author} {\bibfnamefont {A.~A.}\ \bibnamefont
  {Kalachev}}, \bibinfo {author} {\bibfnamefont {S.~L.}\ \bibnamefont
  {Korableva}}, \bibinfo {author} {\bibfnamefont {D.~A.}\ \bibnamefont
  {Sobgayda}}, \ and\ \bibinfo {author} {\bibfnamefont {I.~V.}\ \bibnamefont
  {Zelensky}},\ }\href {\doibase 10.1088/1612-2011/13/1/015202} {\bibfield
  {journal} {\bibinfo  {journal} {Laser Physics Letters}\ }\textbf {\bibinfo
  {volume} {13}},\ \bibinfo {pages} {015202} (\bibinfo {year}
  {2016})}\BibitemShut {NoStop}%
\bibitem [{\citenamefont {Moiseev}\ \emph {et~al.}(2010)\citenamefont
  {Moiseev}, \citenamefont {Andrianov},\ and\ \citenamefont
  {Gubaidullin}}]{Moiseev2010}%
  \BibitemOpen
  \bibfield  {author} {\bibinfo {author} {\bibfnamefont {S.~A.}\ \bibnamefont
  {Moiseev}}, \bibinfo {author} {\bibfnamefont {S.~N.}\ \bibnamefont
  {Andrianov}}, \ and\ \bibinfo {author} {\bibfnamefont {F.~F.}\ \bibnamefont
  {Gubaidullin}},\ }\href {\doibase 10.1103/PhysRevA.82.022311} {\bibfield
  {journal} {\bibinfo  {journal} {Phys. Rev. A}\ }\textbf {\bibinfo {volume}
  {82}},\ \bibinfo {pages} {022311} (\bibinfo {year} {2010})}\BibitemShut
  {NoStop}%
\bibitem [{\citenamefont {Afzelius}\ and\ \citenamefont
  {Simon}(2010)}]{Afzelius2010}%
  \BibitemOpen
  \bibfield  {author} {\bibinfo {author} {\bibfnamefont {M.}~\bibnamefont
  {Afzelius}}\ and\ \bibinfo {author} {\bibfnamefont {C.}~\bibnamefont
  {Simon}},\ }\href {\doibase 10.1103/PhysRevA.82.022310} {\bibfield  {journal}
  {\bibinfo  {journal} {Phys. Rev. A}\ }\textbf {\bibinfo {volume} {82}},\
  \bibinfo {pages} {022310} (\bibinfo {year} {2010})}\BibitemShut {NoStop}%
\bibitem [{\citenamefont {Sabooni}\ \emph {et~al.}(2013)\citenamefont
  {Sabooni}, \citenamefont {Li}, \citenamefont {Kr{\"{o}}ll},\ and\
  \citenamefont {Rippe}}]{2013-Sabooni-PRL}%
  \BibitemOpen
  \bibfield  {author} {\bibinfo {author} {\bibfnamefont {M.}~\bibnamefont
  {Sabooni}}, \bibinfo {author} {\bibfnamefont {Q.}~\bibnamefont {Li}},
  \bibinfo {author} {\bibfnamefont {S.}~\bibnamefont {Kr{\"{o}}ll}}, \ and\
  \bibinfo {author} {\bibfnamefont {L.}~\bibnamefont {Rippe}},\ }\href
  {\doibase 10.1103/PhysRevLett.110.133604} {\bibfield  {journal} {\bibinfo
  {journal} {Phys. Rev. Lett.}\ }\textbf {\bibinfo {volume} {110}},\ \bibinfo
  {pages} {133604} (\bibinfo {year} {2013})}\BibitemShut {NoStop}%
\bibitem [{\citenamefont {Walls}\ and\ \citenamefont {Milburn}(2008)}]{Walls}%
  \BibitemOpen
  \bibfield  {author} {\bibinfo {author} {\bibfnamefont {D.}~\bibnamefont
  {Walls}}\ and\ \bibinfo {author} {\bibfnamefont {G.}~\bibnamefont
  {Milburn}},\ }\href@noop {} {\emph {\bibinfo {title} {Quantum Optics}}},\
  SpringerLink: Springer e-Books\ (\bibinfo  {publisher} {Springer Berlin
  Heidelberg},\ \bibinfo {year} {2008})\BibitemShut {NoStop}%
\bibitem [{\citenamefont {Scully}\ and\ \citenamefont
  {Subairy}(1997)}]{Scully}%
  \BibitemOpen
  \bibfield  {author} {\bibinfo {author} {\bibfnamefont {M.}~\bibnamefont
  {Scully}}\ and\ \bibinfo {author} {\bibfnamefont {M.}~\bibnamefont
  {Subairy}},\ }\href@noop {} {\emph {\bibinfo {title} {Quantum Optics}}}\
  (\bibinfo  {publisher} {Cambridge University Press},\ \bibinfo {year}
  {1997})\BibitemShut {NoStop}%
\bibitem [{\citenamefont {Swanson}\ and\ \citenamefont
  {Macchiarella}(2007)}]{Swanson2007}%
  \BibitemOpen
  \bibfield  {author} {\bibinfo {author} {\bibfnamefont {D.}~\bibnamefont
  {Swanson}}\ and\ \bibinfo {author} {\bibfnamefont {G.}~\bibnamefont
  {Macchiarella}},\ }\href {\doibase 10.1109/MMW.2007.335529} {\bibfield
  {journal} {\bibinfo  {journal} {IEEE Microwave Magazine}\ }\textbf {\bibinfo
  {volume} {8}},\ \bibinfo {pages} {55} (\bibinfo {year} {2007})}\BibitemShut
  {NoStop}%
\bibitem [{\citenamefont {Koziel}\ and\ \citenamefont
  {Ogurtsov}(2011)}]{Koziel2011}%
  \BibitemOpen
  \bibfield  {author} {\bibinfo {author} {\bibfnamefont {S.}~\bibnamefont
  {Koziel}}\ and\ \bibinfo {author} {\bibfnamefont {S.}~\bibnamefont
  {Ogurtsov}},\ }\href@noop {} {\bibfield  {journal} {\bibinfo  {journal}
  {Computational Optimization, Methods and Algorithms}\ ,\ \bibinfo {pages}
  {153}} (\bibinfo {year} {2011})}\BibitemShut {NoStop}%
\bibitem [{\citenamefont {Tamiazzo}\ and\ \citenamefont
  {Macchiarella}(2016)}]{Tamiazzo2017}%
  \BibitemOpen
  \bibfield  {author} {\bibinfo {author} {\bibfnamefont {S.}~\bibnamefont
  {Tamiazzo}}\ and\ \bibinfo {author} {\bibfnamefont {G.}~\bibnamefont
  {Macchiarella}},\ }\href@noop {} {\bibfield  {journal} {\bibinfo  {journal}
  {IEEE Transactions on Microwave Theory and Techniques}\ }\textbf {\bibinfo
  {volume} {65}},\ \bibinfo {pages} {775} (\bibinfo {year} {2016})}\BibitemShut
  {NoStop}%
\bibitem [{\citenamefont {Cheng}(2014)}]{Cheng2014}%
  \BibitemOpen
  \bibfield  {author} {\bibinfo {author} {\bibfnamefont {H.}~\bibnamefont
  {Cheng}},\ }\href@noop {} {\emph {\bibinfo {title} {Integrated Microwave
  Resonator/antenna Structures for Sensor and Filter Applications}}}\ (\bibinfo
   {publisher} {University of Central Florida},\ \bibinfo {year}
  {2014})\BibitemShut {NoStop}%
\bibitem [{\citenamefont {Rosenberg}\ \emph {et~al.}(2013)\citenamefont
  {Rosenberg}, \citenamefont {Salehi}, \citenamefont {Bornemann},\ and\
  \citenamefont {Mehrshahi}}]{Rosenberg2013}%
  \BibitemOpen
  \bibfield  {author} {\bibinfo {author} {\bibfnamefont {U.}~\bibnamefont
  {Rosenberg}}, \bibinfo {author} {\bibfnamefont {M.}~\bibnamefont {Salehi}},
  \bibinfo {author} {\bibfnamefont {J.}~\bibnamefont {Bornemann}}, \ and\
  \bibinfo {author} {\bibfnamefont {E.}~\bibnamefont {Mehrshahi}},\ }\href@noop
  {} {\bibfield  {journal} {\bibinfo  {journal} {IEEE Microwave and Wireless
  Components Letters}\ }\textbf {\bibinfo {volume} {23}},\ \bibinfo {pages}
  {406} (\bibinfo {year} {2013})}\BibitemShut {NoStop}%
\bibitem [{\citenamefont {{\"O}zdemir}\ \emph {et~al.}(2019)\citenamefont
  {{\"O}zdemir}, \citenamefont {Rotter}, \citenamefont {Nori},\ and\
  \citenamefont {Yang}}]{Ozdemir2019parity}%
  \BibitemOpen
  \bibfield  {author} {\bibinfo {author} {\bibfnamefont {{\c{S}}.}~\bibnamefont
  {{\"O}zdemir}}, \bibinfo {author} {\bibfnamefont {S.}~\bibnamefont {Rotter}},
  \bibinfo {author} {\bibfnamefont {F.}~\bibnamefont {Nori}}, \ and\ \bibinfo
  {author} {\bibfnamefont {L.}~\bibnamefont {Yang}},\ }\href {\doibase
  10.1038/s41563-019-0304-9} {\bibfield  {journal} {\bibinfo  {journal} {Nature
  materials}\ ,\ \bibinfo {pages} {1}} (\bibinfo {year} {2019})}\BibitemShut
  {NoStop}%
\bibitem [{\citenamefont {Shakirov}(2007)}]{Shakirov2007}%
  \BibitemOpen
  \bibfield  {author} {\bibinfo {author} {\bibfnamefont {S.}~\bibnamefont
  {Shakirov}},\ }\href@noop {} {\bibfield  {journal} {\bibinfo  {journal}
  {Theoretical and Mathematical Physics}\ }\textbf {\bibinfo {volume} {153}},\
  \bibinfo {pages} {1477} (\bibinfo {year} {2007})}\BibitemShut {NoStop}%
\bibitem [{\citenamefont {Dolotin}\ \emph {et~al.}(2007)\citenamefont
  {Dolotin}, \citenamefont {Morozov},\ and\ \citenamefont
  {Morozov}}]{Dolotin2007}%
  \BibitemOpen
  \bibfield  {author} {\bibinfo {author} {\bibfnamefont {V.~V.}\ \bibnamefont
  {Dolotin}}, \bibinfo {author} {\bibfnamefont {A.}~\bibnamefont {Morozov}}, \
  and\ \bibinfo {author} {\bibfnamefont {A.~D.}\ \bibnamefont {Morozov}},\
  }\href@noop {} {\emph {\bibinfo {title} {Introduction to Non-linear
  Algebra}}}\ (\bibinfo  {publisher} {World Scientific},\ \bibinfo {year}
  {2007})\BibitemShut {NoStop}%
\bibitem [{\citenamefont {Amari}\ \emph {et~al.}(2006)\citenamefont {Amari},
  \citenamefont {LeDrew},\ and\ \citenamefont {Menzel}}]{Amari2006}%
  \BibitemOpen
  \bibfield  {author} {\bibinfo {author} {\bibfnamefont {S.}~\bibnamefont
  {Amari}}, \bibinfo {author} {\bibfnamefont {C.}~\bibnamefont {LeDrew}}, \
  and\ \bibinfo {author} {\bibfnamefont {W.}~\bibnamefont {Menzel}},\
  }\href@noop {} {\bibfield  {journal} {\bibinfo  {journal} {IEEE Transactions
  on Microwave Theory and Techniques}\ }\textbf {\bibinfo {volume} {54}},\
  \bibinfo {pages} {2153} (\bibinfo {year} {2006})}\BibitemShut {NoStop}%
\bibitem [{\citenamefont {Perminov}\ and\ \citenamefont
  {Shakirov}(2009)}]{Perminov2009discriminants}%
  \BibitemOpen
  \bibfield  {author} {\bibinfo {author} {\bibfnamefont {N.}~\bibnamefont
  {Perminov}}\ and\ \bibinfo {author} {\bibfnamefont {S.}~\bibnamefont
  {Shakirov}},\ }\href@noop {} {\bibfield  {journal} {\bibinfo  {journal}
  {arXiv preprint arXiv:0910.5757}\ } (\bibinfo {year} {2009})}\BibitemShut
  {NoStop}%
\bibitem [{\citenamefont {Morozov}\ and\ \citenamefont
  {Shakirov}(2010)}]{Morozov2010new}%
  \BibitemOpen
  \bibfield  {author} {\bibinfo {author} {\bibfnamefont {A.~Y.}\ \bibnamefont
  {Morozov}}\ and\ \bibinfo {author} {\bibfnamefont {S.~R.}\ \bibnamefont
  {Shakirov}},\ }\href {\doibase 10.1007/s11232-010-0044-0} {\bibfield
  {journal} {\bibinfo  {journal} {Theoretical and Mathematical Physics}\
  }\textbf {\bibinfo {volume} {163}},\ \bibinfo {pages} {587} (\bibinfo {year}
  {2010})}\BibitemShut {NoStop}%
\bibitem [{\citenamefont {Karasoulou}(2017)}]{Karasoulou2017algebraic}%
  \BibitemOpen
  \bibfield  {author} {\bibinfo {author} {\bibfnamefont {A.}~\bibnamefont
  {Karasoulou}},\ }\href@noop {} {\emph {\bibinfo {title} {Algebraic
  combinatorics and resultant methods for polynomial system solving}}}\
  (\bibinfo  {publisher} {National and Kapodistrian University of Athens},\
  \bibinfo {year} {2017})\BibitemShut {NoStop}%
\bibitem [{\citenamefont {Reagor}\ \emph {et~al.}(2016)\citenamefont {Reagor},
  \citenamefont {Pfaff}, \citenamefont {Axline}, \citenamefont {Heeres},
  \citenamefont {Ofek}, \citenamefont {Sliwa}, \citenamefont {Holland},
  \citenamefont {Wang}, \citenamefont {Blumoff}, \citenamefont {Chou} \emph
  {et~al.}}]{Reagor2016quantum}%
  \BibitemOpen
  \bibfield  {author} {\bibinfo {author} {\bibfnamefont {M.}~\bibnamefont
  {Reagor}}, \bibinfo {author} {\bibfnamefont {W.}~\bibnamefont {Pfaff}},
  \bibinfo {author} {\bibfnamefont {C.}~\bibnamefont {Axline}}, \bibinfo
  {author} {\bibfnamefont {R.~W.}\ \bibnamefont {Heeres}}, \bibinfo {author}
  {\bibfnamefont {N.}~\bibnamefont {Ofek}}, \bibinfo {author} {\bibfnamefont
  {K.}~\bibnamefont {Sliwa}}, \bibinfo {author} {\bibfnamefont
  {E.}~\bibnamefont {Holland}}, \bibinfo {author} {\bibfnamefont
  {C.}~\bibnamefont {Wang}}, \bibinfo {author} {\bibfnamefont {J.}~\bibnamefont
  {Blumoff}}, \bibinfo {author} {\bibfnamefont {K.}~\bibnamefont {Chou}},
  \emph {et~al.},\ }\href {\doibase 10.1103/PhysRevB.94.014506} {\bibfield
  {journal} {\bibinfo  {journal} {Physical Review B}\ }\textbf {\bibinfo
  {volume} {94}},\ \bibinfo {pages} {014506} (\bibinfo {year}
  {2016})}\BibitemShut {NoStop}%
\bibitem [{\citenamefont {Gorodetsky}\ and\ \citenamefont
  {Ilchenko}(1994)}]{Gorodetsky1994}%
  \BibitemOpen
  \bibfield  {author} {\bibinfo {author} {\bibfnamefont {M.}~\bibnamefont
  {Gorodetsky}}\ and\ \bibinfo {author} {\bibfnamefont {V.}~\bibnamefont
  {Ilchenko}},\ }\href {\doibase 10.1016/0030-4018(94)90603-3} {\bibfield
  {journal} {\bibinfo  {journal} {Optics Communications}\ }\textbf {\bibinfo
  {volume} {113}},\ \bibinfo {pages} {133 } (\bibinfo {year}
  {1994})}\BibitemShut {NoStop}%
\bibitem [{\citenamefont {Brecht}\ \emph {et~al.}(2016)\citenamefont {Brecht},
  \citenamefont {Pfaff}, \citenamefont {Wang}, \citenamefont {Chu},
  \citenamefont {Frunzio}, \citenamefont {Devoret},\ and\ \citenamefont
  {Schoelkopf}}]{Brecht_2016}%
  \BibitemOpen
  \bibfield  {author} {\bibinfo {author} {\bibfnamefont {T.}~\bibnamefont
  {Brecht}}, \bibinfo {author} {\bibfnamefont {W.}~\bibnamefont {Pfaff}},
  \bibinfo {author} {\bibfnamefont {C.}~\bibnamefont {Wang}}, \bibinfo {author}
  {\bibfnamefont {Y.}~\bibnamefont {Chu}}, \bibinfo {author} {\bibfnamefont
  {L.}~\bibnamefont {Frunzio}}, \bibinfo {author} {\bibfnamefont {M.~H.}\
  \bibnamefont {Devoret}}, \ and\ \bibinfo {author} {\bibfnamefont {R.~J.}\
  \bibnamefont {Schoelkopf}},\ }\href {\doibase 10.1038/npjqi.2016.2}
  {\bibfield  {journal} {\bibinfo  {journal} {Npj Quantum Information}\
  }\textbf {\bibinfo {volume} {2}},\ \bibinfo {pages} {16002 EP} (\bibinfo
  {year} {2016})}\BibitemShut {NoStop}%
\bibitem [{\citenamefont {Pierre}\ \emph {et~al.}(2014)\citenamefont {Pierre},
  \citenamefont {Svensson}, \citenamefont {Raman~Sathyamoorthy}, \citenamefont
  {Johansson},\ and\ \citenamefont {Delsing}}]{Pierre2014}%
  \BibitemOpen
  \bibfield  {author} {\bibinfo {author} {\bibfnamefont {M.}~\bibnamefont
  {Pierre}}, \bibinfo {author} {\bibfnamefont {I.-M.}\ \bibnamefont
  {Svensson}}, \bibinfo {author} {\bibfnamefont {S.}~\bibnamefont
  {Raman~Sathyamoorthy}}, \bibinfo {author} {\bibfnamefont {G.}~\bibnamefont
  {Johansson}}, \ and\ \bibinfo {author} {\bibfnamefont {P.}~\bibnamefont
  {Delsing}},\ }\href {\doibase 10.1063/1.4882646} {\bibfield  {journal}
  {\bibinfo  {journal} {Applied Physics Letters}\ }\textbf {\bibinfo {volume}
  {104}},\ \bibinfo {pages} {232604} (\bibinfo {year} {2014})}\BibitemShut
  {NoStop}%
\bibitem [{\citenamefont {Du}\ \emph {et~al.}(2016)\citenamefont {Du},
  \citenamefont {Chau},\ and\ \citenamefont {Zhou}}]{Du2016}%
  \BibitemOpen
  \bibfield  {author} {\bibinfo {author} {\bibfnamefont {H.}~\bibnamefont
  {Du}}, \bibinfo {author} {\bibfnamefont {S.}~\bibnamefont {Chau},
  \bibfnamefont {Fook}}, \ and\ \bibinfo {author} {\bibfnamefont
  {G.}~\bibnamefont {Zhou}},\ }\href {\doibase 10.3390/mi7040069} {\bibfield
  {journal} {\bibinfo  {journal} {Micromachines}\ }\textbf {\bibinfo {volume}
  {7}},\ \bibinfo {pages} {69} (\bibinfo {year} {2016})}\BibitemShut {NoStop}%
\bibitem [{\citenamefont {Melloni}\ \emph {et~al.}(2010)\citenamefont
  {Melloni}, \citenamefont {Canciamilla}, \citenamefont {Ferrari},
  \citenamefont {Morichetti}, \citenamefont {O'Faolain}, \citenamefont
  {Krauss}, \citenamefont {Rue}, \citenamefont {Samarelli},\ and\ \citenamefont
  {Sorel}}]{Melloni2010}%
  \BibitemOpen
  \bibfield  {author} {\bibinfo {author} {\bibfnamefont {A.}~\bibnamefont
  {Melloni}}, \bibinfo {author} {\bibfnamefont {A.}~\bibnamefont
  {Canciamilla}}, \bibinfo {author} {\bibfnamefont {C.}~\bibnamefont
  {Ferrari}}, \bibinfo {author} {\bibfnamefont {F.}~\bibnamefont {Morichetti}},
  \bibinfo {author} {\bibfnamefont {L.}~\bibnamefont {O'Faolain}}, \bibinfo
  {author} {\bibfnamefont {T.~F.}\ \bibnamefont {Krauss}}, \bibinfo {author}
  {\bibfnamefont {R.~D.~L.}\ \bibnamefont {Rue}}, \bibinfo {author}
  {\bibfnamefont {A.}~\bibnamefont {Samarelli}}, \ and\ \bibinfo {author}
  {\bibfnamefont {M.}~\bibnamefont {Sorel}},\ }\href {\doibase
  10.1109/JPHOT.2010.2044989} {\bibfield  {journal} {\bibinfo  {journal} {IEEE
  Photonics Journal}\ }\textbf {\bibinfo {volume} {2}},\ \bibinfo {pages} {181}
  (\bibinfo {year} {2010})}\BibitemShut {NoStop}%
\bibitem [{\citenamefont {Romanenko}\ \emph {et~al.}(2014)\citenamefont
  {Romanenko}, \citenamefont {Grassellino}, \citenamefont {Crawford},
  \citenamefont {Sergatskov},\ and\ \citenamefont {Melnychuk}}]{Romanenko2014}%
  \BibitemOpen
  \bibfield  {author} {\bibinfo {author} {\bibfnamefont {A.}~\bibnamefont
  {Romanenko}}, \bibinfo {author} {\bibfnamefont {A.}~\bibnamefont
  {Grassellino}}, \bibinfo {author} {\bibfnamefont {A.~C.}\ \bibnamefont
  {Crawford}}, \bibinfo {author} {\bibfnamefont {D.~A.}\ \bibnamefont
  {Sergatskov}}, \ and\ \bibinfo {author} {\bibfnamefont {O.}~\bibnamefont
  {Melnychuk}},\ }\href {\doibase 10.1063/1.4903808} {\bibfield  {journal}
  {\bibinfo  {journal} {Applied Physics Letters}\ }\textbf {\bibinfo {volume}
  {105}},\ \bibinfo {pages} {234103} (\bibinfo {year} {2014})}\BibitemShut
  {NoStop}%
\bibitem [{\citenamefont {Huet}\ \emph {et~al.}(2016)\citenamefont {Huet},
  \citenamefont {Rasoloniaina}, \citenamefont {Guillem\'e}, \citenamefont
  {Rochard}, \citenamefont {F\'eron}, \citenamefont {Mortier}, \citenamefont
  {Levenson}, \citenamefont {Bencheikh}, \citenamefont {Yacomotti},\ and\
  \citenamefont {Dumeige}}]{Huet2016}%
  \BibitemOpen
  \bibfield  {author} {\bibinfo {author} {\bibfnamefont {V.}~\bibnamefont
  {Huet}}, \bibinfo {author} {\bibfnamefont {A.}~\bibnamefont {Rasoloniaina}},
  \bibinfo {author} {\bibfnamefont {P.}~\bibnamefont {Guillem\'e}}, \bibinfo
  {author} {\bibfnamefont {P.}~\bibnamefont {Rochard}}, \bibinfo {author}
  {\bibfnamefont {P.}~\bibnamefont {F\'eron}}, \bibinfo {author} {\bibfnamefont
  {M.}~\bibnamefont {Mortier}}, \bibinfo {author} {\bibfnamefont
  {A.}~\bibnamefont {Levenson}}, \bibinfo {author} {\bibfnamefont
  {K.}~\bibnamefont {Bencheikh}}, \bibinfo {author} {\bibfnamefont
  {A.}~\bibnamefont {Yacomotti}}, \ and\ \bibinfo {author} {\bibfnamefont
  {Y.}~\bibnamefont {Dumeige}},\ }\href {\doibase
  10.1103/PhysRevLett.116.133902} {\bibfield  {journal} {\bibinfo  {journal}
  {Phys. Rev. Lett.}\ }\textbf {\bibinfo {volume} {116}},\ \bibinfo {pages}
  {133902} (\bibinfo {year} {2016})}\BibitemShut {NoStop}%
\bibitem [{\citenamefont {Gu}\ \emph {et~al.}(2017)\citenamefont {Gu},
  \citenamefont {Kockum}, \citenamefont {Miranowicz}, \citenamefont {Liu},\
  and\ \citenamefont {Nori}}]{Gu2017}%
  \BibitemOpen
  \bibfield  {author} {\bibinfo {author} {\bibfnamefont {X.}~\bibnamefont
  {Gu}}, \bibinfo {author} {\bibfnamefont {A.~F.}\ \bibnamefont {Kockum}},
  \bibinfo {author} {\bibfnamefont {A.}~\bibnamefont {Miranowicz}}, \bibinfo
  {author} {\bibfnamefont {Y.-x.}\ \bibnamefont {Liu}}, \ and\ \bibinfo
  {author} {\bibfnamefont {F.}~\bibnamefont {Nori}},\ }\href {\doibase
  10.1016/j.physrep.2017.10.002} {\bibfield  {journal} {\bibinfo  {journal}
  {Physics Reports}\ }\textbf {\bibinfo {volume} {718}},\ \bibinfo {pages} {1}
  (\bibinfo {year} {2017})}\BibitemShut {NoStop}%
\end{thebibliography}

%

\end{document}